# Mitigating Systemic Risks in Future Networks
## Use of Utility Functions for Self-Governance

Antonio Manzalini, *Telecom Italia, Strategy - Future Centre*

*Abstract*—This paper elaborates about the potential risk of systemic instabilities in future networks and proposes a methodology to mitigate it. The starting concept is modeling the network as a complex environment (e.g. ecosystem) of resources and associated functional controllers in a continuous and dynamic game of cooperation – competition. Methodology foresees defining and associating utility functions to these controllers and elaborating a global utility function (as a function of the controllers' utility functions) for the overall network. It is conjectured that the optimization of the global utility function ensures network stability and security evaluations. Paper concludes arguing that self-governance (with limited human intervention) is possible provided that proper local, global control rules are coded into these utility functions optimization processes.

## I. INTRODUCTION

TECHNOLOGY trends and socio-economic drivers are steering the evolution of networks towards a connectivity fabric capable of interconnecting huge numbers of interacting heterogeneous nodes. One can easily imagine a scenario, in the near future, where virtual links are dynamically created and destroyed by applications and services to interconnect a very dense environment of processing and storage resources, sensors, actuators, machines, etc.

This scenario will be the socio-economic arena of multiple Players (e.g. Network and Service Providers, OTT, Enterprises, etc.) interacting with each other as in natural ecosystem, for providing any sort of services and data.

In other words, multiple Clouds (figure 1) interconnected through network of networks will create soon a sort of complex system of interconnected nodes, devices, machines.

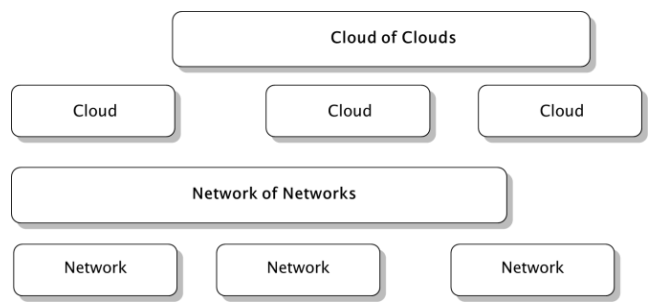

Fig. 1. Multiple clouds across multiple networks

Control and management of this complex environment of resources and services will require, unavoidably, the local-vs-global interaction of multiple controlling systems and methods embedding a certain level of automaticity (in order to ease human operators and mitigate mistakes).

As known from the theory, a complex system is a system consisting of many diverse and autonomous, but interrelated components. Complex systems cannot be easily described by rules and their characteristics are not reducible to one level of description. In fact, complex systems exhibit properties (e.g. self-organization) that emerge from the interaction of their parts and which cannot be predicted from the properties of the single parts.

This means that the increasing level of complexity in future networks will bring new unexpected management challenges and systemic risks. In particular, it is argued that the level of complexity will be soon comparable with the one experienced today in the financial trading market (whose dynamics come from the intertwining of humans operations and automated trading systems).

In order to make an example, in [1] the banking ecosystem is analyzed taking the metaphor of a natural ecosystem as an assembly of species: each of which has feedback mechanisms that would ensure the population's stability only if alone; but the assembly, as a whole, may show sharps transition from overall stability to instability as the number and strength of interactions among species increase. This is very similar to what happens in a financial trading market, coming from the interactions of several Players with automatic trading machineries. In this direction also [2] elaborates about the abrupt system-wide transitions and crashes which may occur out of spontaneous mix of human and fast control machine interactions.

An example is the 'flash crash' of May $6^{th}$ 2010, the second largest point swing in the history of the Dow Jones Industrial Average. For a few minutes, $1 trillion in market value vanished. It has been argued that the 'flash crash' was similar to a complex system transition due to the unexpected coupling of diverse automated trading systems.

Coming back to networks, it is widely recognized that they are strategic assets: in this sense, it is of paramount importance to mitigate the risk of these stability transitions, whose primary effects might be jeopardizing the performance or, in the worst case, creating even a meltdown of a portion of the network.

This paper proposes modeling the network as a complex ensemble (e.g. ecosystem) of resources and functional controllers (solving problems) in a continuous and dynamic game of cooperation – competition. Utility functions are associated to these controllers and a global utility function (in turn a function of the controllers' utility functions) to the network. It is conjectured that the optimization of the network utility function ensures network stability. Paper then concludes arguing that self-governance (with limited human intervention) is possible provided that proper local-vs-global control rules are coded into the utility functions optimization processes.

## II. EXAMPLE OF TODAY NETWORKS INSTABILITIES

Risk of instabilities are already present in today network and cloud infrastructures. This section will make a brief overview, by producing some examples and a selection of the prior art.

### A. Examples and Prior-Art

In a generic communication network, instability of an end-to-end path is a cross-layer issue: in fact, it might depend on the unwanted combination of diverse control mechanisms acting either on the underlying transport network or in the higher layers components (e.g. flow admission control, TCP congestion control and dynamic routing).

For example main arguments for introducing and enhancing flow admission control are essentially derived from the observation that the network otherwise behaves in an inefficient and potentially unstable manner. In fact, even with resources over provisioning, a network without an efficient flow admission control has instability regions that can even lead to congestion collapse in certain configurations.

Another example is congestion control. Currently available mechanisms (like TCP Reno and Vegas) are examples of large distributed control loops designed for ensuring stable congestion control of resources. On the other hand, these mechanisms will be ill-suited, from a stability viewpoint, for future dynamic network, where transients and capacity will be potentially larger [3].

A further example is the instability risk in any dynamically adaptive routing system. Routing instability, which can be (informally) defined as the quick change of network reachability and topology information, has a number of possible origins, including problems with connections, router failures, high levels of congestion, software configuration errors, transient physical and data link problems, and software bugs.

In [4] a simple model of a traditional network traffic dynamics has been presented showing that a phase transition point appears separating the low-traffic phase (with no congestion) from the congestion phase, as the packet creation rate increases. In [5], the previous model has been improved by relaxing the network topology using a random location of routers. This enhanced model has exhibited nontrivial scaling properties close to the critical point, which reproduce some of the observed real Internet features. [6] has discussed a possibility of phase transitions and meta-stability in various types of complex communication networks as well as implication of these phenomena for network performance evaluation and control. Specific cases include connection-oriented networks with dynamic routing, TCP/IP networks under random flow arrivals/departures, and multiservice wireless cellular networks. In [7] the dynamics of traffic over scale-free networks has been investigated. A series of routing of data packets have been proposed, including the local routing strategy, the next-nearest-neighbor routing strategy, and the mixed routing strategy based on local static and dynamic information. Results have indicated the existence of the bi-stable state in the traffic dynamics: specifically the capacity of the network has been quantified by the phase transition from free flow state to congestion state.

[8] has addressed the risk of instabilities in Cloud Computing infrastructures. Paper points out analogies of Cloud Computing infrastructures and complex systems elaborating about the emergence of instabilities due to the unwanted coupling of several reactive controllers.

## III. A METHODOLOGY FOR TAMING INSTABILITIES

This section proposes an example of methodology to mitigate the risk of instabilities in future networks infrastructures.

The approach starts by modeling the network as a complex ensemble (e.g. ecosystem) of resources and functional controllers in cooperation – competition. It is conjectured that the maximization of the network utility function (corresponding to an higher level network controller) ensures network stability. From an implementation viewpoint these controllers can be seen as s/w components (performing distributed computations) pluggable in a lightweight middleware running on top of network equipment [9].

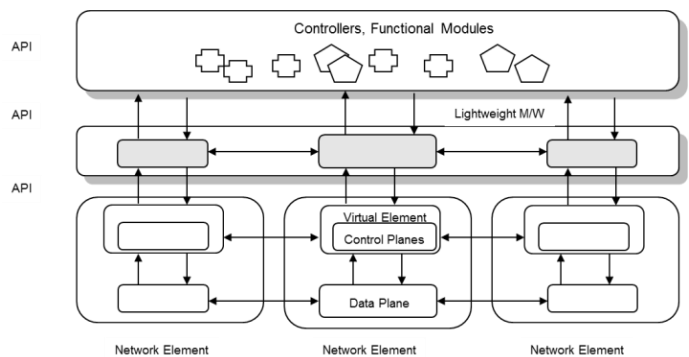

Fig. 2. Example of implementation: a lightweight middleware on top of (current) equipment

This middleware substrate should be as minimal as possible to achieve great scalability, flexibility and evolution without undermining the overall stability e.g. it should offer basic services which are needed for supporting: primitive lifecycle of the network controllers; means of interactions; sets of "fundamental interaction rules" (e.g. pub-sub, reaction-diffusion, excitatory-inhibitory, etc).

While this basic substrate, and its rules, will not have to change (i.e., it will not require re-engineering), the forms under which it manifests itself, as a management framework, can continuously evolve.

In the direction of developing said functional controllers, let's consider the approach reported in [10]. TCP/IP protocol can be seen as an example of optimizer: objective is to maximize the sum of source utilities (as functions of rates) with constraints on resources (figure 3). In fact, each variant of congestion control protocol can be seen as a distributed algorithm maximizing a particular utility function. The exact shape of the utility function can be reverse engineered from the given protocol.

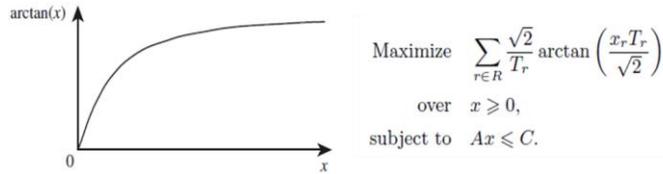

Fig. 3. TCP Utility Function [10]

Similarly, other recent results also show how to reverse engineer Border Gateway Protocols (BGP) as a solution to the Stable Path Problem, and contention-based Medium Access Control (MAC) protocols as a game-theoretic selfish utility maximization [11].

Let's assume to exploit in a network a number of said controllers (i.e. proactive, reactive feedback control loops, methods, etc.) performing network features, or solving certain problems. The problem we wish to highlight is that instability can occur from unintended coupling of independently developed controllers (like in a complex system).

It is conjectured that the problem of ensuring stable network performance can be translated into the optimization problem of an appropriate global utility function(s) for the overall network (or portions of it): this is a functional, i.e. a function of the utility functions associated to the single controllers deployed in the network.

As well known, a utility can be seen a value that represents the desirability of a particular state or set of configurations of the associated system. This is like to say that an utility function can be seen as a function mapping consequences of certain governance decisions into utility values: so maximizing an utility function, $U(\cdot)$, means finding that configuration, $Y_i$, for which we get the maximum utility value: $u_i = U(Y_i)$.

Therefore proposed methodology is based on a three steps approach:
- Decomposing Network problems: this is an analysis required to develop and exploit in network the required set of controllers (e.g. Congestion control, Dynamic routing, Scheduling, Load balancing, Resource Optimization, etc);
- Deriving Controllers' Utility functions: to derive utility functions to be associated to above controllers; each controller is seen as an optimizer whose objective is to maximize its utility, with the related constraints [10];
- Deriving the Network Utility function: to derive the utility functional to be associated to the network, which is an appropriate function of the Controllers' Utility functions.

The task is developing an optimization procedure for maximizing the network utility function, or better to find those network utility values (corresponding to proper configuration of the controllers) so to reach an overall network utility value above a threshold. This is like to say to achieve a stable trade-off for network performance: figure 4 shows a very simple example. Let's consider two controllers: one in charge of optimizing the throughput of the network and another one taking care of cost optimization. On one side the higher is the number of resources allocated, the higher is the throughput of the network, so the utility function of the controller is T=T(n). On the other hand, the cost C=C(n) of the network is a function monotonically increasing with the number of resources allocated. In this very simple example, the overall can be written as U(n) = T(n) – C(n).

The final task is keeping U(n) above a certain threshold (fixed, for example by the SLA).

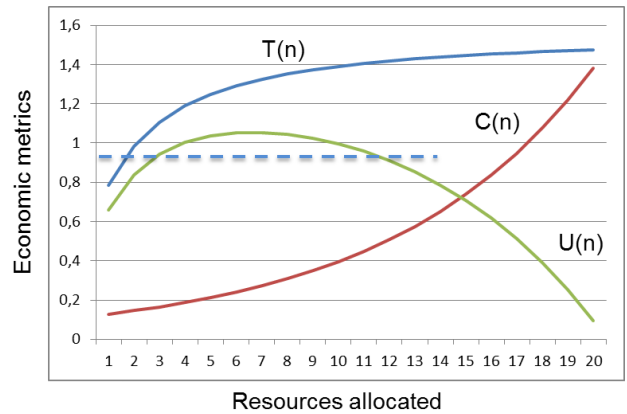

Fig. 4. Example: to achieve a stable trade-off for network performance

It should be noted that this approach has analogies to what happens in natural ecosystems. Consider for example a termite nest: each termite can be seen as a controller, performing certain tasks in the colony. The equilibrium of the ecosystem emerge out of self-governance: single termites contribute, whist optimizing their utility, to optimize the utility of the nest.

### A. Blocks Diagrams descriptions

Imagine a network with M controllers: each controller has a utility function $U_i(\cdot)$, concerning certain performance metrics (figure 5).

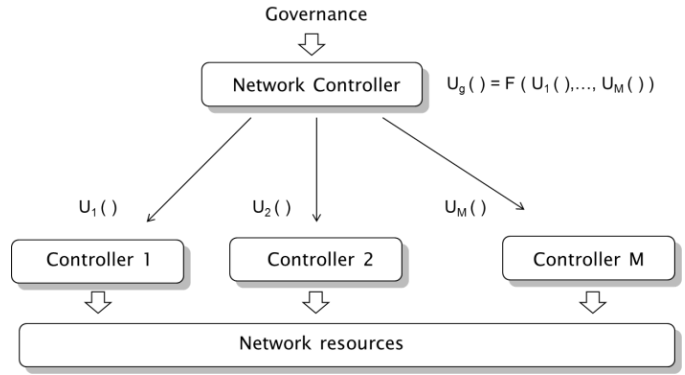

Fig. 5. A network with a set of controllers.

The global utility function of the network is a function F of the utility functions of each controller

$$U_g(\cdot) = F(U_1(\cdot),\ldots,U_M(\cdot)) \quad (1)$$

This is equivalent to say that the network has a global controller (whose utility function is $U_g(\cdot)$ ): this is a sort of orchestrator, which is in charge of configuring the M controllers so to optimize the global utility function.

Maximizing a weighted sum of all utility values is one possible formulation. An approaches may consider multi-objective optimization to characterize the Pareto-optimal tradeoff between the controllers' objectives or game-theory.

Figure 6 and 7 show, respectively, the blocks diagrams of a controller and a network controller.

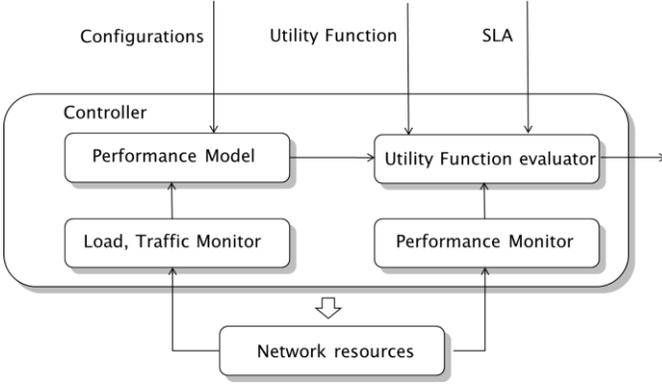

Fig. 7. Block diagram of a Controller

The controller has three main blocks: a monitoring function, a performance model and a utility function evaluator. In particular, the performance model allow adopting specific performance metrics (e.g. throughput as a function of load, traffic and number of resources allocated to a network).

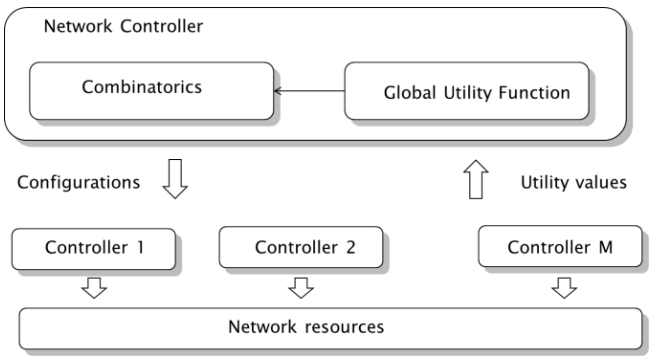

Fig. 7. Block diagram of a Network Controller

In figure 7 the combinatorial search block looks over the space of possible configurations of the parameters of the controllers. This could be done at regular time, upon trigger or to react to changes in the global utility function.

For each set of configurations, then the controllers return the values of their utility functions Ui, which are used to compute the network utility function. When the network controller finds a better configuration vector, it re-starts the configuration process of controllers.

## IV. CONCLUSIONS AND FUTURE WORK

This increasing level of complexity in future networks will bring new unexpected management challenges and systemic risks. In particular, it is argue that such level of complexity will be soon comparable with the one experienced today in the financial market (whose dynamics come from the intertwining of humans operations and automated trading systems). There will be concrete possibilities of having sharps transition from overall stability to instability as the complexity increases.

As networks are strategic assets, it is of paramount importance to mitigate the risk of these stability transitions, whose primary effects might be not only jeopardizing the performance, but even creating even a meltdown of a portion of the network.

In order to start dealing with this problem, paper has proposed a methodology for taming instabilities in complex networks. The starting concept is modeling the network as a complex ensemble (e.g. ecosystem) of resources and functional controllers (solving problems). Utility functions are associated to these controllers and a global utility function (elaborated as a function of the controllers' utility functions) is associated to the network. Paper conjectured that the optimization of the network utility function ensure stability in the network performance.

The conclusion is that networks' stability self-governance is possible provided that we are able of coding and tuning (e.g. in utility functions and functionals) those rules which are governing the delicate interplay of the ensemble of systems composing networks.

Next steps are already concerning the development of a concrete use case to test and demonstrate the feasibility of the proposed methodology. Theoretically the methodology will be also completed by some theorems, still under study.

Moreover this approach (based on optimizing an aggregation function combining single-utility functions from the system components) will be applied for the evaluation of security systems. In fact, the general expression of these functionals allows for an interaction or synergy between the all the components under consideration, and this is just what is needed in security evaluations.

In general, advances in this direction may have far reaching implications and impacts from a socio-economic viewpoint: in this direction, it is of particular interest and relevance to learn the lessons from [12].